\documentclass{PoS}

\title{Exploring Residual Gauge Symmetry Breaking}

\ShortTitle{Exploring Residual Gauge Symmetry Breaking}

\author{Michael Grady\\

        Department of Physics, SUNY at Fredonia, Fredonia NY, 14063 USA\\

        E-mail: \email{grady@fredonia.edu}}

\abstract{Simulations of pure-gauge SU(2) lattice gauge theory are performed in the minimal
Coulomb gauge.  This leaves a residual or remnant gauge symmetry still active which is global 
in three directions but still local in one.  Using averaged fourth-dimension pointing links 
as a spin-like order parameter, the remnant symmetry appears to
undergo spontaneous symmetry breaking at around $\beta = 2.6$.  Both the Binder cumulant 
and the magnetization itself exhibit crossings in this region using lattices up to $20^4$, and 
a susceptibility peak is also observed. Finite size scaling indicates a weak first-order transition. 
The symmetry breaking is also observed to take place in the 
fundamental-adjoint plane,  
and is coincident with the strong first-order transition that exists there at large $\beta _{\rm{adjoint}}$.
This provides confirmation that this phase transition is a symmetry-breaking transition.  A well-known
theorem concerning the instantaneous Coulomb potential has previously proven
that a transition where such a Coulomb-gauge remnant symmetry breaks
is necessarily deconfining.}

\FullConference{XXIVth International Symposium on Lattice Field Theory\\

                July 23-28, 2006\\

                Tucson, Arizona, USA}

\begin{document}
\section{Introduction}
Many gauge fixings leave a remnant gauge symmetry still active, which is global
in at least one dimension.  Once a symmetry is partially global, in the sense that an infinite
number of variables will be affected on an infinite lattice, Elitzur's theorem\cite{elitzur}
no longer applies, and spontaneous symmetry breaking becomes
possible.  Neophytes often wonder why Elitzur's theorem,
which states that a local gauge symmetry cannot break spontaneously, is compatible with the spontaneous
breaking of the local SU(2) gauge symmetry in the standard model.
Although this question is actually quite involved, a quick answer is that
gauge-fixing must be applied in order to study the continuum theory, after which only a global remnant
symmetry survives. It is this remnant gauge symmetry 
which is broken spontaneously by the Higgs particle.  Another place where remnant gauge symmetry apparently takes
place is in continuum quantum electrodynamics\cite{sbqed}.  Here the spontaneous breaking is of the remnant gauge symmetry
left over after fixing to Landau gauge.  It is actually possible to interpret photons as the associated Goldstone
bosons.  

Gauge theories and spin models are closely related, the main difference being 
the local vs. global nature of the symmetry.  In some cases this difference can be removed or at least blurred
through gauge-fixing which at least partially erases the local nature of the symmetry. 
For instance, SU(2) gauge theory
in two dimensions (with open boundary conditions) can be seen to be equivalent to a set of non-interacting
one-dimensional O(4) spin chains if the axial gauge is used.  With all 1-direction links set to
unity, the four-link plaquette interaction is reduced to a two-link spin interaction between the remaining
2-direction links.  In other words, given SU(2) elements 
\begin{equation}
U = a_0 \protect{\bf{1}} + i\sum _{j=1}^{3} a_j \tau _j ,
\end{equation}
the $\rm{tr} UU$ remaining in the action after gauge-fixing 
can be written as $\vec{a_i}\cdot \vec{a_{i+1}}$ with $\vec{a} = (a_0$, $a_1$, $a_2$, $a_3$), 
interpreted as a unit O(4) spin, 
with $i$ a spatial index. This is just a set of
1-d O(4) spin chains.  Another example is the 3-d Z2 lattice gauge theory, dual to the 3-d Ising model.
The duality transformation requires again the use of axial gauge, to make the symmetry groups the same. 

In four dimensions, the differences between spin and gauge theories seem more profound.  Whereas it seems 
every ferromagnetic
spin model in three or more dimensions has a broken-symmetry magnetized phase at low temperature, the non-abelian
gauge theory is expected to have a single confining phase for all couplings. This is quite a different behavior than
any spin model.  In the following I ask if it is truly reasonable that the gauge theory could
act so differently from its cousins.  It would seem a good test would be to apply gauge fixing to eliminate
the Elitzur prohibition on symmetry breaking, and see if the remaining more spin-like
remnant symmetry breaks at weak coupling.  The gauge that makes the 4-d theory look most like a spin
model is minimal Coulomb gauge.  Here, gauge freedom is used to maximize the traces of all links pointing in the 
first three directions, in other words to set gauge elements as close to 
the identity as possible.  At weak coupling this has a profound affect. 
For instance for SU(2), the average trace is about 0.92
at $\beta =$ 2.8, and appears to roughly track the fourth root of the plaquette.  The fourth-direction pointing
links, ignored in the gauge condition, can be analyzed like O(4) spins.  A magnetization can be defined
which is just the average fourth-direction pointing link averaged over each hyperlayer ($N$ separate magnetizations on
an $N^4$ lattice).  The magnetizations transform according to remnant SU(2) symmetries on hyperlayers
perpendicular to the fourth direction.  An SU(2) gauge transformation which is global within the hyperlayers, but
local between, does not affect the gauge condition, so it is the remnant symmetry.  Each O(4) magnetization 
is sensitive to two different remnant symmetries, because tip and tail lie on adjacent hyperlayers. Thus,
if the magnetizations take on an expectation value, the associated remnant symmetry will be spontaneously broken.

This situation is reminiscent of the 2-d case in which the gauge theory splits into non-interacting layers of 
O(4) spins.  However, here the layers are still interacting, though not directly.  Fourth direction links
interact with each other through plaquettes each containing two gauge-fixed links, which due to the gauge fixing
are close to the identity in most cases.
So at weak coupling the largest term in this plaquette is almost always 
a spin-like ferromagnetic dot product between adjacent O(4) spins.
There are of course other terms involving the smaller components of the gauge-fixed links. 
However it seems reasonable to expect that at least at very weak
coupling when the gauge-fixed links get very close to the identity these other interactions 
will become negligible in comparison to the main ferromagnetic terms, and
the gauge theory will begin to act like a layered 3-d O(4) spin model with only weak interlayer interactions.
Magnetization of the 3-d O(4) spin model at low temperature suggests that 
the gauge theory should also magnetize at weak coupling.

The remnant symmetry is intimately connected with confinement. It has been shown previously
that if this symmetry is spontaneously broken, then the instantaneous Coulomb potential 
approaches
a constant at large distances\cite{goz}. Since this potential
is an upper bound for the physical interquark potential\cite{zw}, that too cannot show a linear
increase at large distances in the symmetry-broken phase. Therefore, such a phase is not confining
in the usual sense.  This is also consistent with the 
fact that the O(4) link magnetization order parameter described above is also sensitive to the 
Polyakov loop symmetry, which negates all links in a given direction in a single perpendicular 
hyperlayer.  Applied to the fourth direction, it flips the magnetization.
Thus if these magnetizations take on expectation values, then they will break the Polyakov loop
symmetry, a standard
signal of deconfinement, as well as the remnant symmetry.  
One could simply
use the Polyakov loop as an order parameter, however it is difficult to work with on large
symmetric lattices, because it is numerically small everywhere.  It is also difficult
to determine the expected finite-size scaling behavior due to its nonlocal nature - its definition is
is both lattice-size and boundary-condition dependent.  
The big advantage of the magnetization order parameter is that it is 
the average of a local quantity, so all of the normal methods applied to spin models can be used.
 
\section{Monte Carlo Simulations}

Monte Carlo simulations were run on $12^4$, $16^4$, and $20^4$ lattices with periodic boundary conditions\cite{su2slg},
using 10,000 equilibration sweeps followed by 50,000 measurement sweeps, with quantities
measured after each sweep. The gauge is reset after each sweep, before measurements are taken
(the simulation itself is a standard non-gauge-fixed update).  
Because there are $N$ layers 
there are $N\times 50,000$ values for the magnetization in each run.

\begin{figure}[ht]
                                            \includegraphics[width=2.3in]{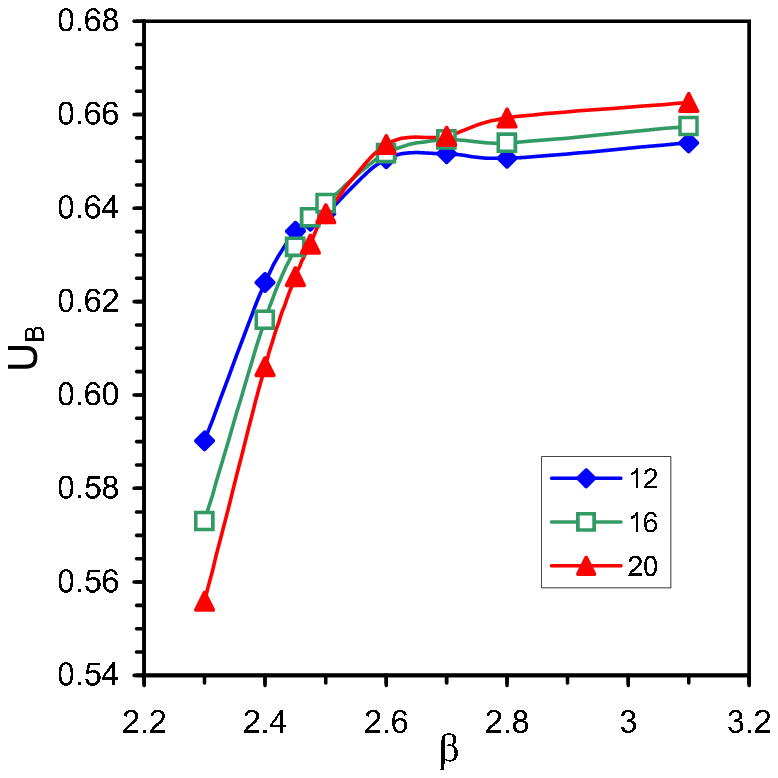}
                                            \includegraphics[width=2.3in]{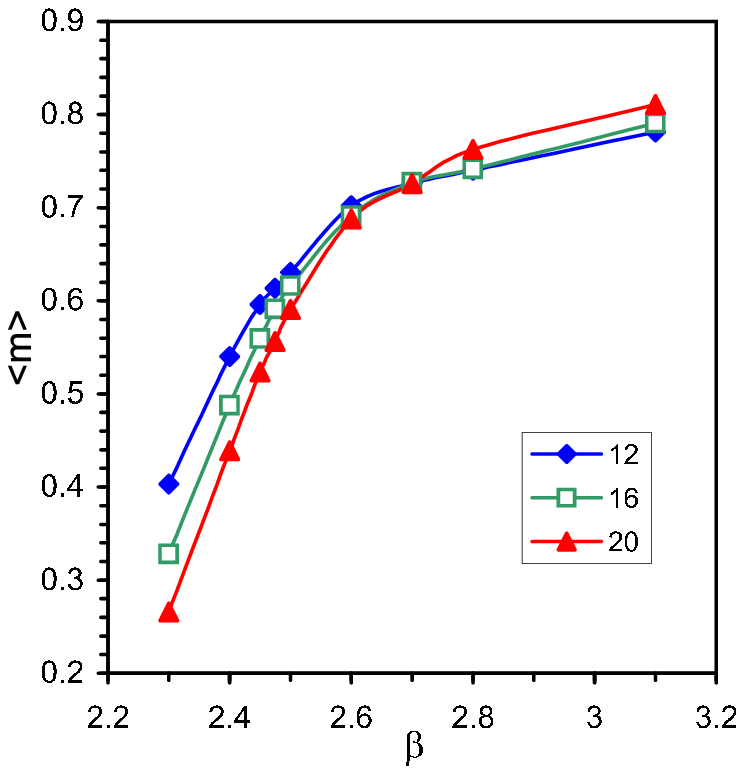}
            \caption{ (a) Binder cumulant 
for $12^4$ (diamonds), $16^4$ (squares), and $20^4$ (triangles) lattices.  Error bars, computed
from binned fluctuations, are about 1/2 the size of points. (b) Layered link magnetizations 
for $12^4$ (diamonds), $16^4$ (squares), and $20^4$ (triangles) lattices.  Error bars are about 1/3 the size of points.}
          \label{fig1}
       \end{figure}
The gold standard for locating the infinite lattice transition point, if there is one, is
to demonstrate crossings of the Binder cumulant as a function of coupling for different lattice sizes.  
For the O(4) order parameter, the Binder cumulant,
defined here as 
$
U_B = 1-<|\vec{m}|^4>/(3<|\vec{m}|^2 >^2 ) ,
$
varies from 1/2 in the full unbroken phase to 2/3 in the fully broken limit\cite{mdop}.
Exactly at the infinite-lattice phase transition it has a nontrivial value in between.
Barring higher order corrections, the $U_B$ curves for all lattice sizes should cross
at this point. When higher order corrections are present, crossings may suffer slight shifts from
each other. 
In Fig.~1(a) a clear crossing is seen in the Binder cumulant around $\beta=2.5$ 
(possibly as high as 2.6). The value of 
$U_B$ at $\beta = 2.8$ and 3.1 on the $20^4$ lattice is more than 10 standard deviations above that for the
$12^4$ lattice, with an even larger
separation in the opposite direction at $\beta = 2.3$.
Fig.~1(b) shows the magnetization, which also exhibits a crossing 
at a slightly higher coupling. Above $\beta=2.7$ the magnetization actually {\em increases} slightly
but significantly with lattice size (five standard deviations separate $12^4$ and $20^4$ values 
at $\beta = 2.8$ and ten at 3.1), 
a strong indication that the magnetization will persist on the
infinite lattice.
\begin{figure}
                      \includegraphics[width=2.5in]{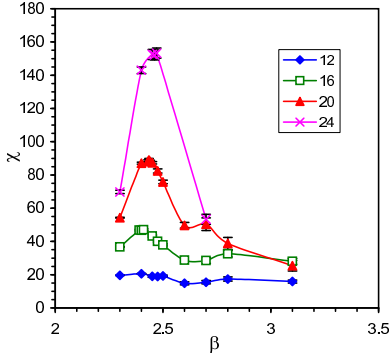}
            \caption{Susceptibility
for $12^4$ (diamonds), $16^4$ (squares), $20^4$ (triangles) and $24^4$ (X's) lattices.}
          \label{fig2}
       \end{figure}
Finally Fig.~2 plots the susceptibility
$\chi = N^3 (<|\vec{m}|^2>-<|\vec{m}|>^2)$,
showing large peaks at the expected location, slightly to the strong coupling side of the 
suspected infinite-lattice critical point.  Peak height is growing rapidly with lattice size. 
The expected finite size scaling of the peak heights is given by $N^{\gamma /\nu }$. 
Using the $16^4$ and $20^4$ peaks,
a value of $\gamma /\nu =2.85 \pm 0.12$ is obtained.  
A few runs were also performed on a $24^4$ lattice. Comparing to the $20^4$ value 
gives $\gamma / \nu = 2.99 \pm 0.15$.
These suggest a weak first-order transition for
which a value of 3, the layer dimensionality, is expected.  Double-peaked histograms at some $\beta$-values
are also suggestive of a first-order transition as is the crossing of magnetization curves, which 
normally doesn't happen for higher-order transitions. 
If the latent heat is small, and is split into $N$ small
mini-jumps as each layer breaks, it could be hidden within normal plaquette fluctuations, which would also
hide it from the specific heat.  

\section{Simulations in Fundamental-Adjoint Plane}
\begin{figure}[hb]
                      \includegraphics[width=2.15in]{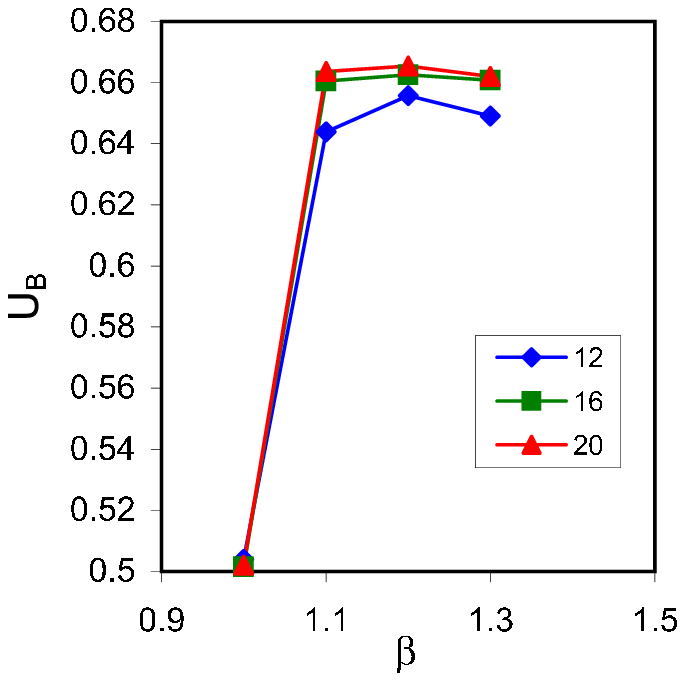}
                      \includegraphics[width=2.25in]{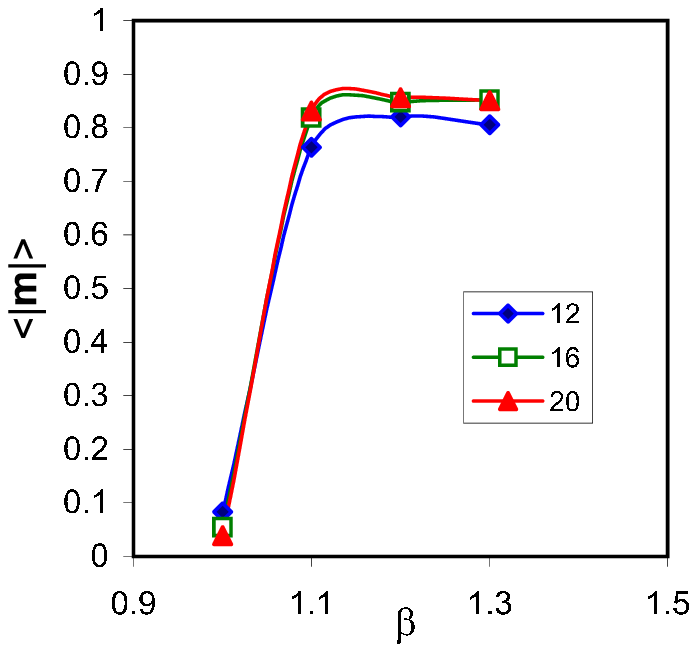}
            \caption{ (a) Binder Cumulant at $\beta _{adjoint} = 1.5$
for $12^4$ (diamonds), $16^4$ (squares), and $20^4$ (triangles) lattices. (b) Magnetization 
for $12^4$ (diamonds), $16^4$ (squares), and $20^4$ (triangles) lattices. }
          \label{fig3}
       \end{figure}

Simulations were also performed in the fundamental-adjoint plane at $\beta _A =1.5$, 
where a strong first-order bulk transition is known to exist,
with latent heat of about 0.26 (plaquette jump).  
It appears clear that the Binder cumulant crossing shown in Fig.~3(a) occurs at the previously-known phase transition 
point (about $\beta =1.04$), as does a crossing in the magnetization seen in Fig.~3(b) (order of points from
different lattices are opposite above and below the transition).
This indicates that this is a symmetry-breaking phase transition, which is consistent with an earlier 
energy analysis\cite{tcp}, but at variance with traditional interpretations.  A symmetry-breaking
transition cannot simply end in a critical point like the liquid-gas transition. It must separate
the entire phase plane into two disjoint phases of different symmetry. 
As explained above, a breaking of the remnant symmetry is necessarily deconfining,
which is consistent with the jump in Polyakov loop Binder cumulant, 
observed to be coincident (Fig.~4).
Therefore, the confining phase must be separated from the weak coupling phase everywhere in the 
fundamental-adjoint plane. 
It is important to realize that the identification of the bulk transition in the fundamental-adjoint plane as
a symmetry-breaking deconfining transition, not unlike that of the Z2 lattice gauge theory,
is sufficient to prove that the continuum limit of zero-temperature SU(2) lattice gauge theory is deconfined.
This is because there is no reason not to take the continuum limit at $\beta _A = 1.5$, $\beta \rightarrow \infty$,
and if the infinite lattice is already deconfined for $\beta > 1.04$ then it is certainly deconfined as
$\beta \rightarrow \infty$.  Unlike on the Wilson axis, there is nothing subtle about this transition.
The latent heat is large and the critical point shows almost no lattice size dependence, so there is
no question about its bulk nature.
In other words, when considering confinement in the continuum limit, 
it is not necessary to determine what happens on the Wilson
axis, though universality would require a phase transition to exist there too. Evidence given above shows
a Wilson-axis transition is present in the magnetization.  However, it is apparently too weak
of a first-order transition there to see directly in the plaquette or specific heat,  
possible reasons for which will be given below.
For this reason the symmetry-breaking nature of the 
stronger transition in the fundamental-adjoint plane is important corroborating evidence for the existence
of a phase transition.
\begin{figure}[ht]
                      \includegraphics[width=2.2in]{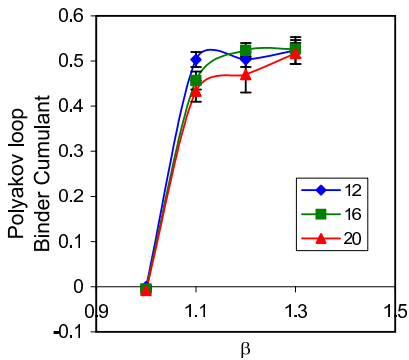}
            \caption{ Binder cumulant of Polyakov loop at $\beta _A = 1.5$ 
for $12^4$ (diamonds), $16^4$ (squares), and $20^4$ (triangles) lattices. 
Polyakov loop itself also jumps, but Binder cumulant is sharper for large lattices.}
          \label{fig4}
       \end{figure}

\section{Discussion and Conclusions}

A layered transition such as the one seen here has some aspects which are three-dimensional, 
because the order parameter is averaged over only the three dimensions within a hyperlayer,
and others which are four because the Hamiltonian is still four-dimensional.  
The transition is more spread out than 
one would expect for lattices of this size, 
because it is the hyperlayer volume that is involved rather than the full volume,
and also due to the binomial expansion factor favoring
partially broken states. For example, there are N ways to choose a single 
broken direction, $N(N-1)/2$ to
choose two, etc. Thus, one needs a higher energy penalty to push the system to a completely broken
or unbroken state. Even on
the $20^4$ lattice the transition is spread from about $\beta =2.3$ to 2.8.
The Polyakov loop transition is seen to occur on the weak side of this transition 
near the point where the last layer becomes broken (around $\beta = 2.75$ on the $20^4$ lattice). 
If the zero-temperature
continuum pure-gauge theory is deconfined, then the deconfinement seen on
asymmetric finite-temperature lattices cannot be a physical transition.  It would be a
modified version of the same transition seen here on symmetric lattices. However, it could
have a different scaling behavior, order, and shifted critical point 
because the finite-temperature lattice is a true 
three-dimensional system; in a similar vein
the three dimensional Ising model with one finite dimension, no matter how large, has the ultimate 
critical behavior of a 2-d Ising model. Also, when fermions are added to the theory, a finite-temperature 
unbreaking of the chiral symmetry is still likely to exist as a physical transition, 
which may have many of the same properties as
a deconfinement transition\cite{tcp,zp,csb,gribov}.

To summarize, the minimal Coulomb gauge allows for the gauge theory to be 
analyzed like a layered magnetic system, with global remnant SU(2) gauge 
symmetries operating separately on each 3-d
hyperlayer.  A link magnetization, which acts as an O(4) spin, magnetizes and breaks the 
symmetry at weak coupling, also breaking the Polyakov loop symmetry. 
A zero-temperature
deconfining phase transition is not expected in a non-abelian theory, but has been 
suggested before\cite{zp,ps}.
The suspected cause of this phase transition is the presence of 
lattice artifacts, similar to the monopoles responsible for
the transition in the U(1) theory.  Some time ago\cite{so3-z2}, a gauge-invariant SO(3)-Z2 monopole was
shown to allow a topologically nontrivial realization of the 
non-abelian Bianchi identity, in a way
analogous to the U(1) monopole in the abelian theory. 
A run was performed on a $12^4$
lattice using minimal Coulomb gauge
at $\beta = 0$ (strong coupling limit) but with SO(3)-Z2 monopoles prohibited
and with plaquettes restricted to be greater than 0.1 (similar to a positive
plaquette constraint, but also avoids plaquettes close to zero which are 
very randomizing for Wilson loops). This run stayed in the broken phase 
of the layered link magnetization
with a distribution similar to the Wilson-action simulation
at $\beta = 3.1$. 
Nevertheless, this action produces an interquark potential similar to that seen with the
Wilson action at $\beta=2.85$,
so it may be a practical way to avoid the artifacts and access continuum physics on
reasonable-sized lattices. 

A possible reason for the existence of a weak first-order transition in virtually 
all gauge theories is that when a symmetry breaks on the infinite lattice, an ergodic restriction
occurs which prevents tunneling to other sectors. This would appear to result in a sudden change in entropy.
If this entropy change is extensive, then a latent heat  = $T \Delta s$ exists. At first glance, it
would appear that the number of symmetries breaking, $N$ on an $N^4$ lattice, would not be sufficient, in that
the associated entropy jump would scale like $N$. However there is an additional gauge freedom in the minimal
Coulomb gauge due to exceptional configurations. If the sum of the six links pointing in the one through
three directions touching a site is zero, then a gauge transformation there will not affect the gauge condition.
Although extremely rare, the number of such sites in a gauge configuration scales with volume. In counting
the number of ergodically prohibited gauge transformations away from a given configuration, one must include 
combinations of such ``exceptional'' gauge transformations with the symmetry-violating ones, giving a
small but non-zero jump in the specific
entropy 
which will be seen as a latent heat associated with a first-order phase transition.

\end{document}